\documentstyle[11pt,newpasp,twoside,epsf]{article} 
\def\etal{{\it et al. }}
\begin{document}
\title{Spectroscopy of Brown Dwarf Candidates in the NGC~1333 Molecular Cloud}
\author{Bruce Wilking, Ayman Mikhail, Glenn Carlson}
\affil{Department of Physics and Astronomy, University of Missouri-St. Louis, 8001 Natural Bridge Road, St. Louis, MO 63121}

\author{Michael Meyer}
\affil{Steward Observatory, The University of Arizona, Tucson, AZ 85721}

\author{Thomas Greene} 
\affil{NASA/Ames Research Center, M.S. 245-6, Moffett Field, CA 94035}

\begin{abstract}
We present an analysis of low-resolution infrared spectra for 25
brown dwarf candidates in the NGC~1333 molecular cloud.  Candidates
were chosen on the basis of their association with the high column
density cloud core, and near-infrared fluxes and colors.
We compare the depths of water vapor
absorption bands in our candidate objects with a grid of dwarf,
subgiant, and giant standards to determine spectral types which are
independent of reddening.  These data are used to derive effective
temperatures and bolometric luminosities which, when combined with
theoretical tracks and isochrones for pre-main sequence objects,
enable us to estimate masses and ages.  Depending on the models considered, a total of 9 to 20 brown dwarfs
are identified with a median of age of $<$1 Myr.
\end{abstract}

\section{Introduction}

The NGC~1333 reflection nebula and its associated dark cloud L~1450 are
part of a chain of molecular clouds in the Perseus region (Sargent
1979; Loren 1976).  Analysis of Hipparcos data suggests a distance of 300
pc (de Zeeuw, Hoogerwerf, \& de Bruijne 1999; Belikov \etal 2002).  The observations of
emission-line stars and Herbig-Haro objects first established NGC~1333 as
an active region of star formation (e.g., Herbig 1974).
Surveys of the cloud at near-infrared
wavelengths have revealed a large population of low mass stars (Strom,
Vrba, \& Strom 1976; Aspin, Sandell, \& Russell 1994; Lada, Alves, \& Lada 1996; Wilking \etal 2002).
The young stellar objects are concentrated into a
northern and southern cluster, each with about 70 members.
Identification of the lowest mass objects in the NGC~1333 cloud, as well as
estimates for their ages and masses, requires infrared spectroscopy.
Ultimately, the population of very low mass objects will be used to define
the shape of the mass function and determine the number and total mass of
brown dwarfs in the young cluster.  

\section{Source Selection and Observations}

Brown dwarf candidates were selected by comparing their positions in a near-infrared color-magnitude
diagram 
with theoretical models for pre-main sequence objects
by D'Antona \& Mazzitelli (1997, 1998, hereafter DM97).  An age of 1 Myr was assumed
(Lada \etal 1996).
Near-infrared surveys of the northern cluster (Wilking \etal 2002) and 
the southern cluster (Aspin \etal 1994)
were used to construct the K vs. (H-K) diagram.
The majority of the 25 candidates lie in a region
of the diagram where M$<$0.1 M$_{\sun}$,
A$_v<$10 mag, and K$<$14 mag.  

Infrared spectroscopic observations were accomplished using the 3-m
NASA Infrared Telescope Facility at Mauna Kea, Hawaii in 2000
November 10-13.  The observations were aided by an Internet2 link between UM-St. Louis 
and Mauna Kea.  Spectra in the K band were obtained for
25 brown dwarf candidates, eight M giants, and eight M subgiants using the
256 x 256 InSb facility infrared camera (NSFCAM) with the HKL
grism and a 0.3\arcsec\ pixel$^{-1}$ scale.
The 0.6\arcsec\ slit provided a resolution of R = $\lambda$/$\Delta\lambda\sim$300 
over the 2.0 $\mu$m--2.5 $\mu$m band.

\section{Results}

\subsection{Spectral Classification Using a Water Vapor Index}

In the K-band, broad water vapor absorption bands centered at 1.9 $\mu$m and 2.5 $\mu$m
provide the best means for deriving spectral types for faint M stars.  These bands are not
only sensitive to temperature for cool stars (Jones \etal 1994) but also are well-resolved
in low resolution spectra.  Water vapor indices have been used to derive spectral
types for very low mass young stars in the $\rho$ Ophiuchi cloud (Wilking, Greene, \& Meyer 1999,
hereafter WGM99; Cushing, Tokunaga, \& Kobayashi 2000),
IC~348 (Najita, Teide, \& Carr
2000), the Orion Nebula cluster (Lucas \etal 2001), 
and the Taurus cloud (Itoh, Tamura, \& Tokunaga
2002).
In WGM99, we defined a Q index that was independent of reddening as
\begin{equation}
Q = (F1/F2)(F3/F2)^{1.22}
\end{equation}
where F1, F2, and F3 are the average fluxes in narrow bands
covering 2.07--2.13 $\mu$m, 2.2695--2.2875 $\mu$m, and 2.40--2.50 $\mu$m,
respectively.

A linear fit to a plot of the Q index vs. the optically-determined
M spectral type for dwarf standards is shown in Fig. 1 for stars from M0.5V--M9V
and yields the following relation
\begin{equation}
MV~subclass = (-20.15\pm0.48) \times Q + (18.49\pm0.55)
\end{equation}
with a correlation coefficient of r=0.98 for a sample of 12 measurements.
\begin{figure}
\plotfiddle{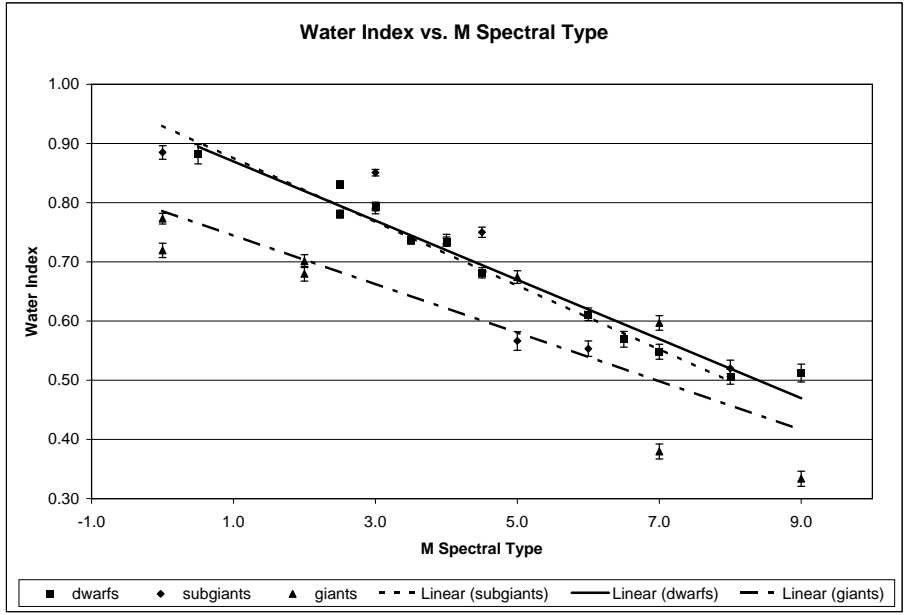}{3.0truein}{0.0}{50}{50}{-200}{-36}
\caption{The Q index vs. M spectral type for dwarf standards (squares), subgiants (diamonds), and
giants (triangles).  Linear fits to the dwarf and subgiant data are nearly
identical.}
\end{figure}

\subsection{The Water Index as a Function of Surface Gravity}

A sample of eight M subgiants were selected for infrared spectroscopic
observations from the MBM~12 and IC~348 young stellar
populations.
The median ages of these populations are estimated to
be 2 Myr.
As shown in Fig. 1, the linear fit to the subgiant sample is
consistent with the dwarf relation and there is no apparent sensitivity of the
Q index to surface gravities in the dwarf to subgiant range.
This is in agreement with the findings of Jones \etal (1995) who, in comparing infrared
synthetic spectra, saw little dependence on the depths of the water
lines with surface gravity between log(g)=4.0 to 5.0.
The Q values for the 8 giants observed show a much greater scatter
from a linear fit.
In our sample, dwarfs always display deeper water vapor absorption than giants
for a given spectral type.  
However, the Q index for
giants can be smaller than that of dwarfs 
due to enhanced absorption by CO in the
2.40--2.50 $\mu$m band.  
Hence, the Q index is not a useful measure of water vapor absorption 
for objects with the surface gravities of giant stars.

\subsection{Spectral Types}

Twenty-four brown dwarf candidates displayed late-type photospheres with absorption due to
water vapor and CO characteristic of high surface gravity.  
Since the surface gravities
of objects in this sample are likely to resemble those of subgiants,
spectral types were derived using Eqn. 2 which is appropriate for both subgiants
and dwarfs.
Spectral classifications using the water vapor index relation
reveals spectral types between M2.4 and M8.3 with typical
uncertainties  of $\pm$0.5 subclass.  Seventeen of 24
candidates displayed spectral types $\ge$M6.0.  
The visual extinctions implied by these spectral types are typically 2-3 mag and all less than
10 mag.
The majority of the candidates appear to be young objects of very low mass.  The projection of
this sample on the high column density core (A$_v>$10 mag) coupled with the relatively
low visual extinctions of the candidates minimizes the chance that some objects are
background M stars.  The detection of x-ray emission from 15 of the 24 candidates by Getman \etal (2002),
including
3 tentative detections, argues in favor of the youth of this sample, as x-ray luminosity
and stellar activity are
known to decline with age for K and M stars (Fleming, Schmitt, \& Giampapa 1995).

\subsection{Masses and Ages of the Brown Dwarf Candidates}

We followed the procedure outlined in WGM99 to
estimate the masses and ages of the brown dwarf candidates.
Effective temperatures computed for each brown dwarf candidate have 
typical
uncertainties of $\pm$95 K.
However, systematic effects could affect the derived temperatures.
The presence of moderate veiling in 3 objects could
lead us to overestimate their effective temperatures by $\sim$150 K.
The assumption of dwarf, rather than subgiant, surface gravities could
also systematically affect our results.
For spectral types of M2 and later, giant star temperatures are
warmer than dwarfs by 300--500 K for stars of the same spectral type
(e.g., Fig. 4 in Itoh \etal 2002).  Hence by assuming a dwarf surface gravity,
we may be underestimating the effective temperature of an M2 subgiant by $\sim$150 K
and of an M6--M9 subgiant by $\sim$250 K.

In Fig. 2, we have plotted our brown dwarf candidates on Hertzsprung-Russell diagrams
overlaid with the theoretical tracks and isochrones from the models of
of DM97 and Burrows et al. (1997).  An H-R diagram using the Baraffe \etal (1998) models is 
presented in Wilking \etal (2002).
For the DM97 models, mass
estimates range from $<$0.02 M$_{\sun}$ to 0.25 M$_{\sun}$, with 16 objects at or
below the hydrogen-burning limit.  The median age for the sample is 0.3 Myr.  If a temperature
scale intermediate between dwarfs and giants is used, then these objects should be shifted to higher
temperatures by 250 K and only 9 would remain as brown dwarfs.  
The Burrows \etal models yield lower masses and
younger ages with 23 objects at or
below the hydrogen-burning limit and 20 of these remaining as brown dwarfs if 
shifted to higher temperatures by 250 K.  The median age for the sample implied by the
Burrows \etal models is 0.1 Myr.
\begin{figure}
\plottwo{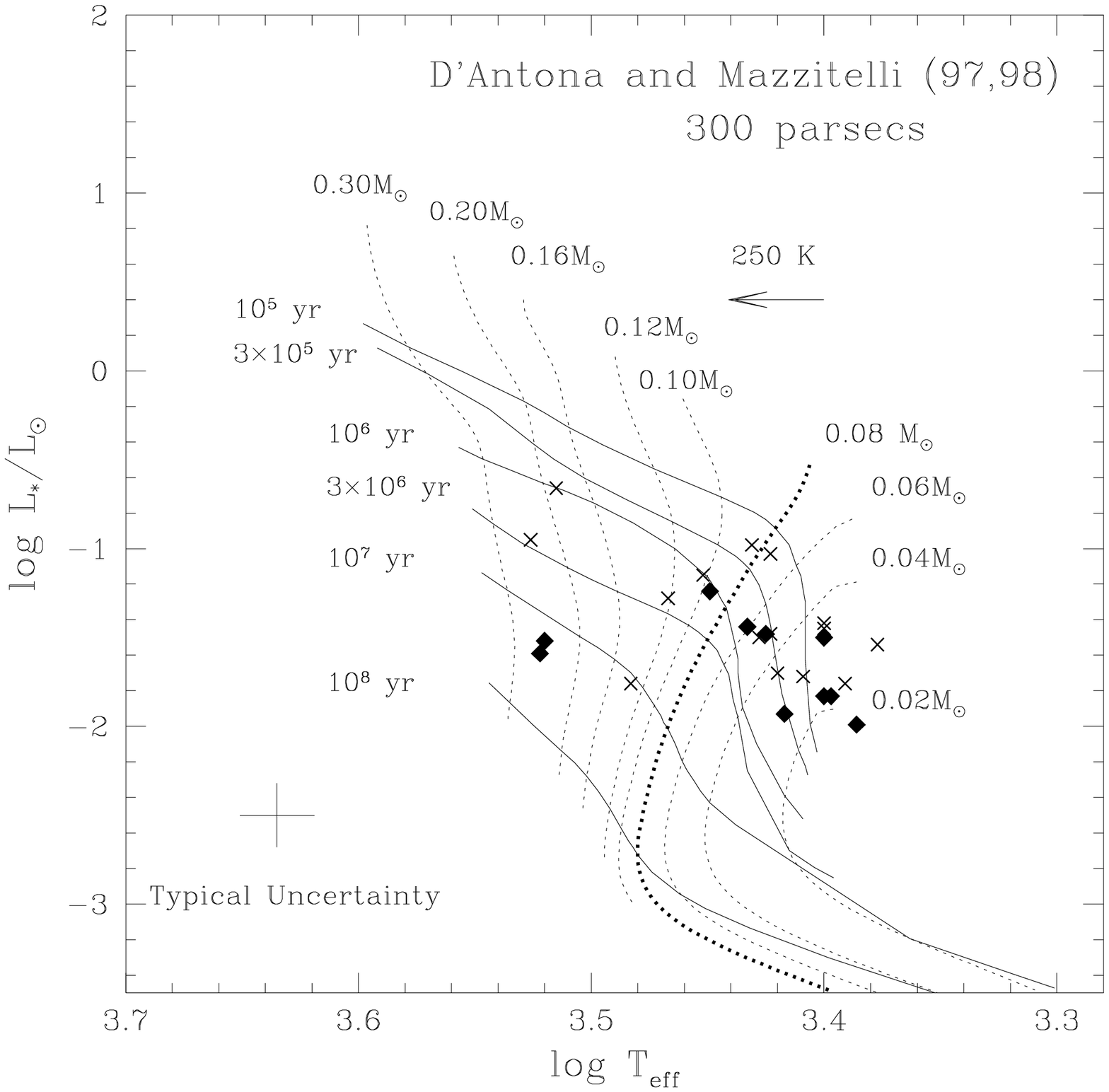}{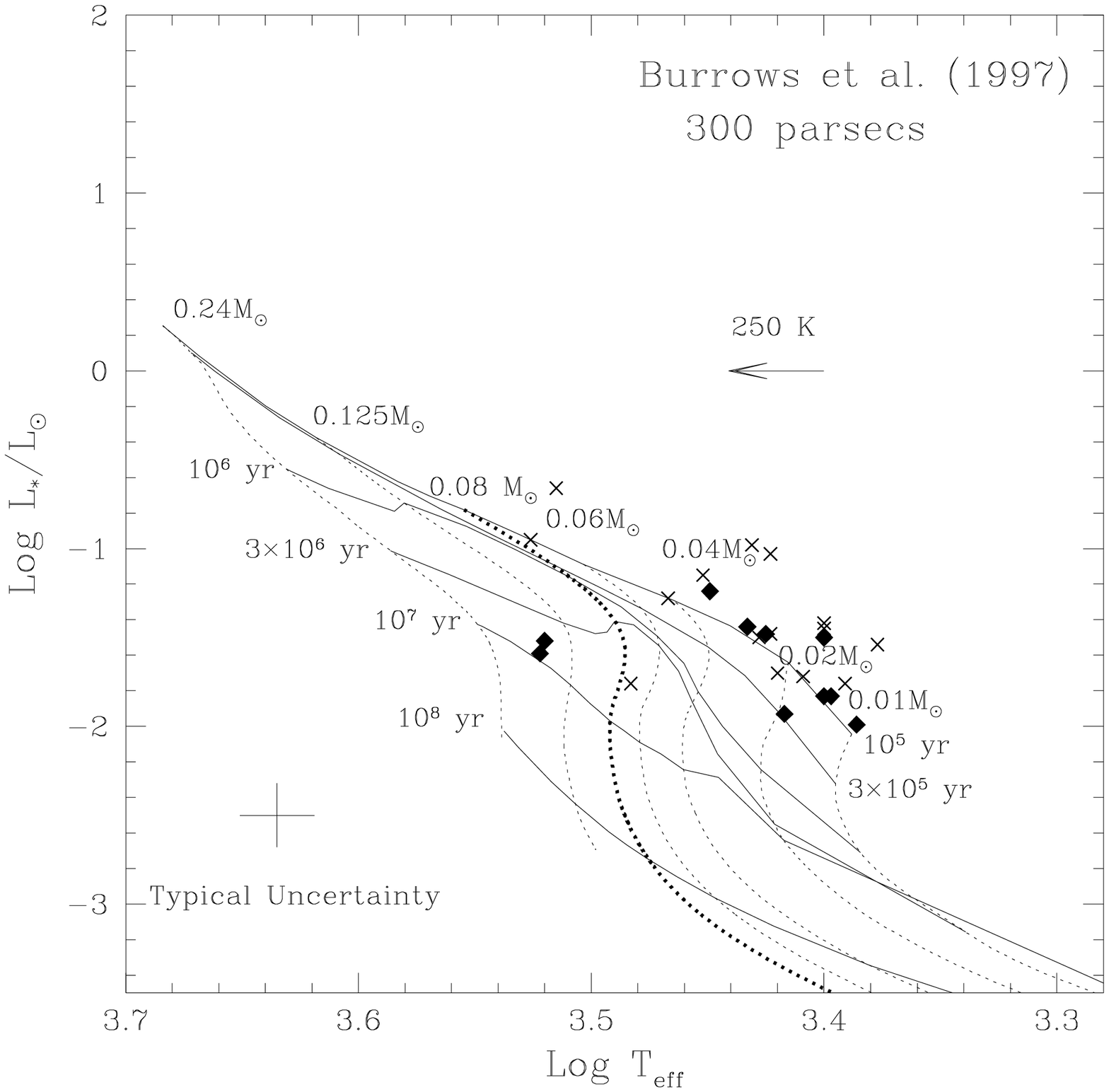}
\caption{H-R diagrams for the brown dwarf candidates using the DM97 models (left panel)
and the Burrows \etal models (right panel).  `X's mark the positions
of x-ray emitting candidates.}
\end{figure}

\subsection{Surface Gravity Estimates}

Surface gravities (g=GM/R$^2$) can be estimated for the brown dwarf
candidates using the DM97 mass estimates and radii
derived from the luminosity and effective temperature.
Values range from log(g)=3.0 to 4.5 in cgs units with a median value
of 3.3.  These estimates assume a temperature scale, intrinsic colors,
and bolometric corrections derived from dwarf standards and only increase
under the assumption of giant properties. Therefore, the surface gravities of our objects
more closely resemble dwarfs than giants, implying that our use of a dwarf temperature
scale has led to an underestimation of the temperature for M6--M9 stars of
$<$250 K.  When taken along with the results of the two sets of models, we
conclude that at least 9 of the candidates can be identified as true brown dwarfs.

\end{document}